\documentstyle[12pt]{article}
\newcommand{\beq}{\begin{equation}}
\newcommand{\eeq}{\end{equation}}
\newcommand{\bdm}{\begin{displaymath}}
\newcommand{\edm}{\end{displaymath}}
\newcommand{\bea}{\begin{eqnarray}}
\newcommand{\eea}{\end{eqnarray}}

\newcommand{\th}{\theta}

\newcommand{\zb}{\bar{z}}

\newcommand{\ib}{\bar{\imath}}
\newcommand{\jb}{\bar{\jmath}}
\newcommand{\nn}{\nonumber}
\newcommand{\wc}{{\cal W}}
\begin{document}
\setcounter{page}{0}
\topmargin 0pt
\oddsidemargin 5mm
\renewcommand{\thefootnote}{\fnsymbol{footnote}}
\newpage
\setcounter{page}{0}
\begin{titlepage}
\begin{flushright}
USP-IFQSC/TH/93-12
\end{flushright}
\vspace{0.5cm}
\begin{center}
{\large {\bf Affine Toda Field Theory in the Presence of  Reflecting
Boundaries}} \\
\vspace{1.8cm}
{\large Andreas Fring}\footnote{ Supported by FAPESP - Brasil.
             Address after $1^{st}$ of October 1993: Department of Physics,
Univ
   ersity  College of Swansea. Swansea SA2 8PP, UK.}
{\large and Roland K\"oberle\footnote{ Supported in part by CNPq-Brasil.}}
\footnote{ FRING@BR.ANSP.USP.IFQSC  and  ROLAND@IFQSC.ANSP.BR } \\
\vspace{0.5cm}
{\em Universidade de S\~ao Paulo, \\
Caixa Postal 369, CEP 13560 S\~ao Carlos-SP, Brasil\\}
\vspace{3cm}
\renewcommand{\thefootnote}{\arabic{footnote}}
\setcounter{footnote}{0}
\begin{abstract}
{ We show that the ``boundary crossing-unitarity equation" recently proposed
by Ghoshal and Zamolodchikov is a consequence of the boundary bootstrap
equation for the S-matrix and the wall-bootstrap equation. We solve this
set of equations for all affine Toda theories related to simply laced Lie
algebras, obtaining explicit formulas for the W-matrix which encodes the
scattering of a particle with the boundary in the ground state. For each theory
there are two solutions to these equations, related by CDD-ambiguities, each
giving rise to different kind of physics. }
\end{abstract}
\vspace{.3cm}
\centerline{August 1993}
 \end{center}
\end{titlepage}
\newpage
\section{Introduction}

The central element in a quantum field theory, on-shell, is its scattering
matrix, which relates the asymptotic in- and out states. As shown in the
seminal
    paper by  Zamolodchikov and Zamolodchikov\cite{ZZ} integrable field
theories
    in 1+1 dimensions posses an  n-particle S-matrix, which factorises into
2-particle S-matrices, which can then be determined exactly.
\par
Several problems in quantum mechanics, like dissipative systems\cite{Leg}
can be understood as a quantum field theory in the presence of boundaries
\cite{Callan}. In particular, one might get some more insight into the space
of boundary states in open string theory \cite{Witten}. The general scheme of
this approach was initiated by
Cherednik \cite{Ch1}, who studied the S-matrix describing the scattering of
part
   icles off a wall. In particular the question of how
to find and solve the factorisation (Yang-Baxter)
equation in the presence of a reflecting wall has been answered. Recent
research
    has added bootstrap\cite{FK} and crossing/unitarity equations\cite{GZ},
whic
   h complete the set of equations necessary to compute the 2-particle S-matrix
   in a theory with boundaries.
\par
For a theory to remain integrable the boundary conditions should  maintain a
suf
   ficient number of conservation laws. For example in
conformal field theories \cite{Cardy}, integrability demands  the boundary
condi
   tions not to introduce any length scale. This allows conformal invariance to
   be imposed, requiring now momentum conservation parallel to the boundary to
b
   e hold.
As discussed in \cite{GZ} a breaking of conformal invariance has to respect the
   conservation laws, selecting in this way {\em integrable } boundary
condition
   s. Formally the integrals of motion are found to be generalizations
of the Hamiltonian to higher spins
\beq
H_s \; = \; \int_{- \infty}^{0} dx \left( T^{\zb \zb}_{s+1} + T^{\zb z}_{s-1}
          -T^{z z}_{s+1} - \bar{T}^{\zb z}_{s-1} \right) + \phi_B(y)
\eeq
where  the {\em boundary perturbation }
$\phi_B(y)$ denotes some local field. The quantities $H_s$ do not
depend on $y$, if
\beq
\frac{d \phi_B(y)}{dy} \; =\;  T^{\zb \zb}_{s+1} + T^{\zb z}_{s-1}
          -T^{z z}_{s+1} - \bar{T}^{\zb z}_{s-1} |_{x=0} \;\;\; .
\label{eq: intcond}
\eeq
In the present paper we shall be concerned
with affine Toda field theories (ATFTs) \cite{MOP} and we will assume the
existe
   nce of
such boundary conditions\footnote{We hope to report elsewhere
on a proper investigation of this question.}. ATFTs constitute an
important example of completely integrable models, since they are explicit
Lagrangian versions of integrable deformations of conformal field theories.
The breaking of the conformal symmetry simply corresponds to an affinisation
of the Lie algebra {\bf g}, underlying the conformal invariant theory.
Many other features can be expressed as well very neatly in terms of Lie
algebraic quantities, which makes them interesting objects to study even
from a purely mathematical point of view.
\par

Our presentations falls into two main sections. In the following we review
some of the features of the scattering matrix of ATFTs in order to
establish our notation and to formulate the equations needed, in particular
(\ref{eq: bootsp}). Section three
is the central part of our manuscript. We derive a {\em pair-bootstrap}
equation
    for the S-matrix and show that, together with the wall-bootstrap equation,
t
   hey imply the   boundary crossing-unitarity eqution.
We then  present the full set
of constraining equations for the W-matrix, which encodes scattering due to the
   wall, and solve it for all simply laced ATFTs. Finally we state our
conclusio
   ns.

\section{The S-matrix of Affine Toda Field Theory}

In the original formulation by Cherednik \cite{Ch1} of a scattering theory
which includes boundaries, a distinction is made between the scattering
matrices encoding the process before or after some particles have hit the
boundary. In general these matrices turn out to be different, but are related
by
some relatively simple relations \cite{Ch1,Ch2,Sk,MN,KSA,DeVega}.
Assuming diagonality of all scattering matrices involved, this distinction
becomes obsolete \cite{FK} and one solely has to deal with one type of
 S-matrix. Presuming further the existence of some suitable boundary
conditions, in the sense of (\ref{eq: intcond}), the theory will again be
purely elastic. The S-matrix will then  be the product of two factors:
one encodes scattering off the boundary and the other coincides with
the S-matrix in the absence of boundaries. It is this second factor, which
we review in this section.

 As the result of a sequence of
investigations \cite{AFZ,KS,Muss,Dest,KM,BCDS,PD,FO}  the S-matrix of
affine Toda field theories related to simply laced Lie algebras has been found
to take on a very compact form \cite{PD,FO}
\footnote{ For the non-simply laced case several candidates exist for an
S-matrix. Refer to \cite{FK2} and references therein.} .
 Adopting the notation of \cite{FLO,FO} it reads
\beq
S_{ij}(\th) = \prod_{q=a}^{b} \left\{ 2q - \frac{ c(i) + c(j)}{2}
\right\}_{\th}^{- \lambda_i \cdot \sigma^q \gamma_j } \;\;\; .
\label{eq: smat}
\eeq
Here $\th$ denotes the relative rapidities, where $\th_i$ parameterises
as usual the momenta $p_i = m_i ( \cosh \th_i , \sinh \th_i ) $.
The limits are taken to be $a = \frac{c(i) +1}{2}$ and $b= \frac{h-1}{2} +
\frac{c(i) + c(\ib)}{4} $. $\{ \}_{\th}$ is a building block
comprised out of sinh-functions, i.e. $\{ x\}_{\th} = [ x ]_{\th}/ [x]_{-\th},
[x]_{\th} = < x+1>_{\th} <x-1>_{\th} / < x+1-B>_{\th} <x-1+B>_{\th} $ and
$< x>_{\th} = \sinh\frac{1}{2} \left( \th + \frac{i \pi x}{h}\right)$.
$B(\beta)$ is a real function between 0 and 2 incorporating the coupling
constant dependence of the Lagrangian field theory, which is assumed to be
real in the following. In finding the solution
of functional equations it is useful to employ the following equivalent
integral representation for each block
\beq
\{ x \}_{\theta} = \exp \left( \int_{0}^{\infty} \frac{ dt} {t \sinh t}
\; f_{x,B}(t) \;\; \sinh \frac{ \theta t} {i \pi} \right) \label{eq: intrep}
\eeq
where
\beq
f_{x,B}(t) = 8 \sinh \frac{t B}{2 h} \sinh \frac{t}{h} \left( 1 - \frac{B}{2}
\right) \sinh t  \left( 1 - \frac{x}{h} \right) .
\eeq
In each affine Toda theory related to a simple Lie algebra {\bf g} there are
$r \equiv$ rank of {\bf g} scalar fields, describing
r particles, which can either be associated with a fundamental
weight $ \lambda_i$ or a simple root $ \alpha_i$. The map $\sigma$, emerging
in the exponent of (\ref{eq: smat}), denotes the Coxeter element of the Weyl
group, being a product of reflections in a complete set of simple roots.
In order to define a ``universal" element $\sigma$, that is an element which
can be written in a general form, independent of what Lie algebra {\bf g} one
is concerned with, one associates signs
$+$ and $-$ to the vertices of the Dynkin diagrams and defines a function
$ c(i) = \pm 1 $ which identifies them. Then the two elements $\sigma_{\pm}$
of the Weyl group consisting of reflections related to simple roots
associated with $\pm$-signs can be used to define uniquely and unambiguously
a particular Coxeter element $ \sigma := \sigma_- \sigma_+$ for any Lie
algebra {\bf g}. The Coxeter element splits all the roots  into r
disjoint orbits containing each h, being the Coxeter number,
elements. The roots $\gamma_i = c(i) \alpha_i$ lie all in distinct orbits
$\Omega_i$ \cite{FLO} and one can therefore alternatively associate a whole
orbit to a particle.
\par
It can be shown \cite{FO} that $S_{ij}(\th)$ (\ref{eq: smat}) is a meromorphic
function which satisfies the usual crossing and unitarity relations demanded
from the two particle S-matrix
\beq
S_{ij}(\th) S_{ij}(-\th) = 1 \qquad \hbox{and} \qquad
S_{i \jb}(\th) = S_{ij} ( i \pi - \th) \;\;   .  \label{eq: crossuns}
\eeq
In the case the three point coupling $C_{ijk}$ is non-zero, $S_{ij}(\th)$
will posses an odd order pole with positive residue due to the propagation
of a bound state particle $k$. Then (\ref{eq: smat}) satisfies the so-called
bootstrap equation formulated originally in \cite{ZZ} in
order to find constraining equations, which pose an equivalent to the
Yang-Baxter equation \cite{YB} for diagonal S-matrices
\beq
S_{li}\left( \th + i \eta_i \right) \;S_{lj}\left( \th + i \eta_j \right) \;
S_{lk}\left( \th + i \eta_k \right) \; = \; 1   \;\; .\label{eq: boots}
\eeq
Here the ``fusing angles" $\eta_t$, for $t=i,j,k$, are given by
$ \eta_t = - \frac{\pi}{h} \left( 2 \xi(t) + \frac{1 - c(t)}{2} \right)$
and are related to the fusing rule, conjectured in  \cite{PD}, which decides
whether the coupling constant $C_{ijk}$ is vanishing or not. It states that
only if there exist equivalence classes of integers $(\xi(i), \xi(j), \xi(k) )$
and $(\xi'(i), \xi'(j), \xi'(k) )$ satisfying
\beq
\sum_{t=i,j,k} \sigma^{\xi(t)} \gamma_t =0 \qquad \hbox{and} \qquad
\sum_{t=i,j,k} \sigma^{\xi'(t)} \gamma_t =0  \;\; , \label{eq: fuseroot}
\eeq
then the three-point coupling $C_{ijk}$ will be non-zero. The two classes
correspond to the only two inequivalent solutions which are related via
$\xi'(t) = - \xi(t) + \frac{c(t) -1}{2} $ \cite{FO}. Classically this
rule simply corresponds to the non-vanishing of the commutator involving
two stepoperators of the Lie algebra ${\bf g}$ $\left[ E_{\sigma^{\xi(i)}
\gamma_{i} } ,  E_{\sigma^{\xi(j)}\gamma_{j} } \right]$ \cite{FLO}, whereas
quantum mechanically it becomes equivalent to the bootstrap equation
(\ref{eq: boots}).
\par
Making use of the identity $\gamma_i = ( \sigma_- - \sigma_+) \lambda_i$
we can re-express (\ref{eq: fuseroot}) in terms of fundamental weights
\beq
\sum_{t=i,j,k} \sigma^{-\xi'(t)} \lambda_t =0 \qquad \hbox{and} \qquad
\sum_{t=i,j,k} \sigma^{-\xi(t)} \lambda_t =0 \label{eq: fusew}
\eeq
which will be of importance below.
\par
Further we require a particular combination of the boostrap equations in the
next section. Selecting the three equations in (\ref{eq: boots}) for
$l = i,j,k$ and substituting them into each other we derive the {\em
pair-bootst
   rap } equation
\beq
S_{kk}\left( \th + i 2 \eta_k \right) =
S_{ii}\left( \th + i 2\eta_i \right) \;S_{jj}\left( \th + i 2\eta_j \right) \;
S_{ij}^2\left( \th + i \eta_i  + i \eta_j \right)  \;\; .\label{eq: bootsp}
\eeq
This means having a pair of two particles $i$ and $j$, possessing rapidities
which will give rise to a bound state, it will be equivalent to scattering
either the two pairs of particles or the two bound states against each other.
We depict this situation in figure 1.
\par
This equation posses as well a geometrical formulation, in the same manner as
(\ref{eq: boots}) can be re-expressed geometrically in terms of
(\ref{eq: fuseroot}) and (\ref{eq: fusew}).

 To see this we  consider
$$
S_{ii}(\th + 2 i \eta_i)  =  \prod_{q=1}^{h} \left\{ 2q - c(i)
\right\}_{ \th + i2 \eta_{i}}^{- \frac{1}{2} \lambda_i\cdot\sigma^q\gamma_{i} }
 \; = \; \prod_{q=1}^{h}  \left( \frac{ [2q - 4 \xi(i) -1 ]_{2 \th} }
{ [-2q - 4 \xi(i) + 2c(i) - 1 ]_{2 \th} } \right)^{- \frac{1}{2} \lambda_i
\cdot \sigma^q \gamma_{i} }
$$
where we have employed the relation $ \{ x \}_{ \th + \frac{  \pi i y}{h} }
= \frac{ [ x + y ]_{ \th}} { [ y -x ]_{ \th}}$ .
Shifting now in the numerator and denominator by
$$
q \rightarrow q + 2\xi(i) \qquad \hbox{and} \qquad  q \rightarrow q - 2 \xi(i)
+
c(i) -1
$$
respectively, gives
\beq
S_{ii}(\th + 2i \eta_i) \; = \; \prod_{q=1}^{h}
\frac{ [2q-1 ]_{ \th}^{-\frac{1}{2}\lambda_i\cdot\sigma^{q+2\xi(i)}\gamma_{i}}}
 { [-2q+1 ]_{\th}^{-\frac{1}{2}\lambda_i\cdot\sigma^{q + 2\xi'(i)\gamma_{i}}}}
 \;\;  . \label{eq: shw}
\eeq
Performing similar manipulations on the expression for the square of the
S-matrix yields
\beq
S^2_{ij}(\th + i \eta_i + i \eta_j) \; = \; \prod_{q=1}^{h} \frac
{ [2q-1 ]_{ \th}^{-\lambda_i\cdot\sigma^{q +\xi(i)+ \xi(j)}
\gamma_{j} }}{ [-2q+1]_{\th}^{- \lambda_i\cdot\sigma^{q +
 \xi'(i) + \xi'(j)} \gamma_{j} }} \;\;  . \label{eq: shs}
\eeq
Employing  now the first equation in (\ref{eq: fuseroot}) and  the second in
(\ref{eq: fusew}) we derive the relation
\beq
\lambda_i \cdot \sigma^{q + 2 \xi(i)} \gamma_i + \lambda_j \cdot \sigma^{q + 2
\xi(j)} \gamma_j + 2 \lambda_i \cdot \sigma^{q +  \xi(i) + \xi(j) } \gamma_j
= \lambda_k \cdot \sigma^{q + 2 \xi(k)} \gamma_k \; .
\label{eq: multp}
\eeq
A similar equation can be obtained for the second solution for the fusing rule
involving $\xi'(t)$ instead of $\xi(t)$.
Assembling now this results, that is multiplying the expression (\ref{eq: shw})
for $i$ and $j$ with (\ref{eq: shs}), the power of the building blocks will
take on half the left hand side of (\ref{eq: multp}), such that
(\ref{eq: bootsp}) is satisfied.
\par
The scattering matrix exhibits the usual CDD-ambiguity \cite{CDD}
 $ S_{ij}(\th) \rightarrow \psi_{ij}(\th)$ $ S_{ij}(\th)$, being determined
only
    up to the factor $\psi_{ij}(\th)$. Equations (\ref{eq: crossuns}) will pose
   the following
constraint on this function
\beq
\psi_{ij}\left( \th + \frac{i \pi}{2} \right) \psi_{i\jb} \left( \th - \frac{i
\pi}{2} \right) \;= \; 1      \;\;\;  ,      \label{eq: crossuncdd}
\eeq
which is solved by any function of the form
\beq
\psi_{ij}(\th) = \prod_x \frac{ < x>_{\th} <h-x>_{\th}}{ <
x>_{-\th}<h-x>_{-\th}
   }
\;\; .
\eeq
Further restrictions come from the bootstrap equations
\beq
\psi_{li}\left( \th + i \eta_i \right) \;\psi_{lj}\left( \th + i \eta_j \right)
\; \psi_{lk} \left( \th + i \eta_k \right) \; = \; 1 \label{eq: bootcdd}
\eeq
such that the scattering matrix is fixed up to the function $\psi_{ij}$,
which we require not to introduce new poles in the physical sheet, since these
w
   ould have to participate in the bootstrap.
\par
It should be noted that a scattering matrix can be thought of as resulting
from the braiding of two operators $Z_i(\th)$ and $Z_j(\th)$, associated
to particle $i$ and $j$, respectively, in the Zamolodchikov algebra
\cite{ZZ}
\beq
Z_i(\th) Z_j(\th)   \; = \; S_{ij}(\th) Z_j(\th) Z_i(\th)
\label{eq: zamalg}
\eeq
of which an explicit representation for the matrix (\ref{eq: smat}) has been
found in \cite{CD}. Assuming the existence of some vacuum $ | 0 \rangle$
these operators can be used to construct the Hilbert space by successive
action of $Z_i^{\dag} (\th)$ on this state, that is the Hilbert space is
spanned
by the operators $ \prod\limits_{i=1}^{h} Z_i^{\dag} (\th) | 0 \rangle$. Hence
by interpreting the left hand side of (\ref{eq: zamalg}) as an in-state and
the two operators on the right hand side as an out-state, the S-matrix acquires
its original sense as defining  the superposition of out-states as in-states.
\par

\section{The W-matrix of Affine Toda Field Theory}
In the presence of reflecting boundaries the introduction of an additional
matrix $W_{1 \dots N}^{1' \dots N'} (\th) $, which encodes the scattering
of particles off the boundary, is required. We assume the existence of
some conserved quantities such that this matrix factorises, into one-particle
amplitudes $W^{i '}_{i}(\th)$, in a similar fashion as the S-matrix.
Furthermore
we presume its diagonality, such that particle $Z_i(\th)$ does not change its
quantum numbers while scattering from the wall, but solely reverses its
momentum. Then extending the algebra (\ref{eq: zamalg}) by an operator
$Z_w(0)$, representing the wall, the one-particle reflection amplitude
$W_i(\th)$ results from
\beq
Z_i(\th) Z_w(0) \; = \; W_i(\th) Z_i(-\th) Z_w(0)  \;\; . \label{eq: defw}
\eeq
The operator $ Z_w(0)$ is thought to define the ground state of
the Hilbert space in the presence of the boundary, i.e. $ |W \rangle :=
 Z_w(0) | 0 \rangle$, such that now the superposition of in-states in terms
of out-states is governed by a product of S- and W-matrices.
\par
In analogy to a derivation of the first equation in (\ref{eq: crossuns}),
that is applying twice (\ref{eq: zamalg}), from its very definition
(\ref{eq: defw}), the equivalent unitarity equation for $W_i$ results to be
\beq
W_i(\th) W_i(-\th) \; = \; 1   \;\;\; . \label{eq: uniw}
\eeq
As in the theory without boundaries, the associativity of the algebra
(\ref{eq: zamalg}) gives rise to some factorization equations \cite{Ch1,FK},
which however for diagonal S-and W-matrices contain no information.
Instead we require the analogue to the
bootstrap equation (\ref{eq: boots}) in the presence of a reflecting boundary,
which was derived in \cite{FK}. Relating again the fusing angles to the
integer powers which occur in the fusing rule, this equation reads now
\beq
W_k( \th + i \eta_{\bar{k}}) \;= \; W_i( \th + i \eta_{i} ) W_j( \th
+i\eta_{j})
S_{ij} ( 2 \th + i \eta_{i} +i\eta_{j} ) \;\; .
\label{eq:bootw}
\eeq

As was pointed out by Ghoshal and Zamolodchikov \cite{GZ} the equivalent
to the second equation in (\ref{eq: crossuns}), namely the crossing relation,
is far less obvious. We shall provide an alternative derivation of their
crossing unitarity equation and demonstrating that in fact this equation
is implied by the wall bootstrap equation. Shifting $\th$ by $i \pi$ in
(\ref{eq:bootw}) and subsequently multiplying the resulting equation
by (\ref{eq:bootw}) yields
\bea
W_k( \th + i \eta_{\bar{k}}) W_k( \th + i \eta_{\bar{k}} + i \pi) &= &
W_i( \th + i \eta_{i} ) W_i( \th + i \eta_{i} + i \pi )
W_j( \th +i\eta_{j}) \nonumber \\
 & &  \times  W_j( \th +i\eta_{j} + i \pi)S_{ij}^2 ( 2 \th + i \eta_{i}
 +i\eta_{j} ) \;\; . \nonumber
\eea
Comparision of this equation with (\ref{eq: bootsp}) shows, that it is solved
if
\beq
W_i( \th) W_{\ib} (\th +i \pi) \; =\;  S_{ii} (2 \th)
\label{eq: crossunw}
\eeq
is true. This is precisely the ``cross-unitarity equation"
originally derived in \cite{GZ}, employing the fact that  the same correlation
f
   unction can be  viewed in two ways related by interchanging space and time
co
   ordinates.
\par
{}From the symmetry of the S-matrix in its indices we obtain with
(\ref{eq: crossunw}) that the W-matrix for particles and anti-particles
coincide
\beq
W_{\ib}( \th) \; = \; W_{i} (\th)\;\; .  \label{eq: antip}
\eeq
Furthermore from the $2 \pi i $-periodicity  of the S-matrix we  obtain
by means of (\ref{eq: crossunw}) that W is $2 \pi i$-periodic in $\th$ too.
Using this facts we may rewrite (\ref{eq: crossunw}) as
\beq
W_{\ib}\left(\th +\frac{i \pi}{2}\right) W_i\left(\th -\frac{i \pi}{2}\right)
S_{i \ib} (2\th)
\; = \; 1 \;\; .   \label{eq: crossunw2}
\eeq

Equation (\ref{eq: crossunw}) can be given the following meaning. Consider the
p
   article
$i$ living in the half-space $-\infty<x\leq 0$ delimited by
a wall at $x=0$. Let it scatter with rapidity $\th_1 = \th$ against the wall,
this process being
described by $W_i( \th)$. Now the other factor of the left hand side
of equ.(\ref{eq: crossunw}) $W_{\ib}(\th +i \pi) $ represents the scattering
of the anti-particle $\ib $, but with rapidity increased by $i\pi$, which
is the same as particle $i$ at rapidity $\th_2 = -\th$ penetrating the
boundary. This corresponds to a process in a different physical region, namely
scattering from the wall at $x=0$ by a particle living in the half-space
$0\geq x <\infty$. Both processes are shown in fig. 2. Now $W_i(\theta)$
is not relativistically invariant, but describes the scattering off the
wall in the prefered frame in which the wall is at rest.
We may now look at the scattering of two particles $i$ with
rapidities $\theta_1,\theta_2$ in this particular
frame with the same asymptotic configuration as the one described by
the left hand side of equ.(\ref{eq: crossunw}) and also illustrated in fig. 2.
We have $\theta_1=+\theta,\;\theta_2=-\theta$ and the corresponding $S$-matrix
is $S_{ii}(2\theta)$, which is exactly the right hand side of equ.(\ref{eq:
cros
   sunw}). This equation tells us therefore, that an observer far away from the
   wall cannot tell whether two incoming particles with opposite rapidities
scat
   ter against each other at $x=0$ or whether they have been
reflected by a double-sided mirror, which has been placed at the origin.
\par
One might now worry whether equations (\ref{eq: crossunw}) and (\ref{eq:bootw})
are compatible for all possible angles. We observe that taking
$j = \ib $ in the wall bootstrap equation and shifting $\th$ by $ -i
\frac{\eta_i}{2} - i \frac{\eta_{\ib}}{2}$, the right hand side of
(\ref{eq:bootw}) takes on exactly the same form as (\ref{eq: crossunw2}) since
mod$2 \pi i$  $ \eta_i -\eta_{\ib} = \pm \pi$. However in that case we
have $ \xi(i) -\xi(j) = \pm \frac{h}{2} + \frac{c(i) - c(j)}{4}$, which
is precisely the power of the Coxeter number required to relate the
representatives of the orbit $\Omega_i$  and $\Omega_{\ib} $ , related to
particles and antiparticles, respectively \cite{FO}. Hence this case does
not pose a problem, since the fusing rule will never give rise to these angles.
\par
In \cite{FK} we solely had equations (\ref{eq: uniw}) and (\ref{eq:bootw})
at our disposal, which we solved for some specific Toda models. It was
demonstrated there that it is in principle possible to find solutions of
this system of equations, but due to the lack of (\ref{eq: crossunw}) it is
a rather involved procedure. As demonstrated above, together with the
homogeneous bootstrap equation, one equation can be derived from the other.
Hence instead of solving (\ref{eq:bootw}) we now solve  first
the ``crossing-unitarity" equation (\ref{eq: crossunw}).
Employing the standard technique of Fourier transforms, we obtain
an integral representation for the W-matrix:
\beq
W_i(\th) =  \exp \left( \frac{-1}{2 \pi}   \;\;\int d \th' \;
 \frac{1}{ \cosh( \th -\th')} \ln\,S_{i \ib}(2 \th') \right)  \; . \label{eq:
in
   tw}
\eeq
For the theories in mind we know that the S-matrix will always be of the
form $ \prod\limits_x \{ x \}_{\th}$ and we therefore obtain that the W-matrix
will acquire the same form
\beq
W( \theta ) \; =  \; \prod_x {\cal W}_x ( \theta ) \;\; , \label{eq: wfac}
\eeq
where the blocks ${\cal W}_x ( \theta )$ are in one-to-one correspondence
to the ones in S. Using the integral representation (\ref{eq: intrep})
we can carry out the $\th'$-integration in (\ref{eq: intw}) and obtain
\beq
{\cal W}_x ( \theta ) \; = \; \frac{ w_{1-x} ( \theta)  w_{-1-x}(\theta) } {
w_{
   1-x - B} ( \theta)  w_{-1-x + B} ( \theta) } \;\; ,\label{eq: subblock}
\eeq
where the subblocks $ w_x(\th)$ are given by
\beq
w_x(\th)  = \frac{ < \frac{x-h}{2} >_{\th} } { < \frac{x-h}{2} >_{-\th} } \; .
\eeq
One easily verifies the relations
\bea
 w_x ( \theta ) \;\;  w_x (- \theta ) & = & 1   \\
 w_{x-2h} ( \theta ) \;  w_{-x} ( \theta ) &=& 1 \\
 w_x ( 0 ) =  w_{-h} ( \theta )  &=& 1 \\
 w_x \left( \theta + \frac{i y \pi}{2h}\right)\;
 w_x \left( \theta - \frac{i y \pi}{2h}\right)  &=&
 w_{x+y} ( \theta )  \;  w_{x-y} ( \theta )  \\
 w_x \left( \theta + \frac{i \pi}{2}\right)\;
 w_x \left( \theta - \frac{i \pi}{2}\right)  &=&
\frac{ <x >_{2 \th} } { <x >_{-2 \th} }
\eea
from which we deduce the unitarity
\beq
{\cal W}_x ( \theta ) {\cal W}_x ( -\theta ) \; = \; 1
\eeq
and the crossing-unitarity relation
\beq
{\cal W}_x \left( \th +\frac{i \pi}{2} \right) {\cal W}_x \left(\th -\frac{i
 \pi}{2} \right) \{ x \}_{2 \th} \; = \; 1  \label{eq: crossunx}
\eeq
for each block in (\ref{eq: wfac}). Furthermore we derive
\bea
\wc_{ x + h} ( \th) \wc_{x-h} (\th) \{ x \}_{2 \th} &= & 1   \label{eq: ssx} \\
\wc_{2h - x } ( \th) \wc_{x} (\th)  &= & 1 \label{eq: merom} \\
\wc_{x} (\th + 2 \pi i)  & =& \wc_{x} (\th) \;\; .
\eea
The zeros and poles of each block ${\cal W}_x ( \theta )$ are simple and lie
on the imaginary $\th$-axis. The poles lie mod$2 \pi i$ at
\beq
\th_{\pm} = \frac{ \pm 1 - x -h}{ 2h} i \pi \qquad \hbox{and} \qquad
\th_{\pm}^{B} = \frac{ \pm B \mp 1 + x +h}{ 2h} i \pi
\eeq
whereas the zeros are situated at
\beq
^0 \th_{\pm} = \frac{ \pm 1 + x +h}{ 2h} i \pi \qquad \hbox{and} \qquad
^0\th_{\pm}^{B} = \frac{ \pm 1 \mp B - x -h}{ 2h} i \pi  \;\; .
\eeq
Notice  that for $ 0 < x < h$, $\th_{\pm}$ will never lie in the physical
sheet,
    but the coupling constant dependent pole can now, on the contrary to
the case of the S-matrix, move inside $ 0 <\hbox{ Im }\th < i \pi$.
\par
Analogous to the $S$-matrix, we also have a CDD-ambiguity related to the
$W$-matrix:  $ W_i(\th) \rightarrow W_i(\th) \psi_i( \th) $, where
$\psi_i(\th)$ satisfies the homogeneous equations  (\ref{eq: crossuncdd}),
with $ j = \jb = 0$ indicating the ground state, and  (\ref{eq: bootcdd}),
with $l=0$. Thus the CDD-ambiguity for the S- and W-matrices turn out to be
restricted by the same equations. However in the latter case it can be
obtained in simple way.
Whereas for the S-matrix a shift of $\th$ by $i \pi$ gives simply rise to
essentially the inverse, the same shift for the W-matrix gives rise to a new
function. We observe that if $\wc_x(\th)$ solves equation (\ref{eq: crossunx}),
then $\wc_{x+2h} (\th) = \wc_x(\th + i \pi)$ will be a solution as well,
which from (\ref{eq: ssx}) is evidently not simply the inverse. It is easy
to verify that the function $\psi_x(\th)$ which relates these two solutions
\beq
\wc_x(\th + i \pi) \; =\; \wc_{ x + 2h}(\th)\;=\; \wc_x(\th) \psi_x(\th) \; ,
\eeq
indeed blockwise solves (\ref{eq: crossuncdd}) and (\ref{eq: bootcdd}).
Introduc
   ing a product over these
blocks we observe that if $ \wc_t(\th)$ for $t=i,j,k$ are solutions of
(\ref{eq: crossunw}) and (\ref{eq:bootw}) then $ \wc_t(\th + i \pi)$
 will solve them likewise and hence the function
\beq
\psi_i(\th) \;=\; \frac{ \wc_i(\th + i \pi)} { \wc_i(\th) }
\eeq
obeys the CDD-constraints (\ref{eq: crossuncdd}) and (\ref{eq: bootcdd}).
\par
Drawing a close analogy to the S-matrix (\ref{eq: smat}) of affine Toda theory
we might now conjecture its W-matrix to be of the form
\beq
W_i(\th) = \prod_{q=a}^{h-1+a} \left( {\cal W}_{2q - \frac{c(i)+c(\ib)}{2}
 }(\th)\right) ^{- \lambda_i \cdot \sigma^q \gamma_{\ib} } \;\; .
\label{eq: wmat}
\eeq
In the following we shall verify that this function, apart from the
bootstrap equation, indeed satisfies the general consistency requirement
expected from it.
\par
Since each block $W_x(\th) $ individually satisfies the unitarity relation
(\ref{eq: crossunx}), $W_i(\th)$ will do so likewise. In order to satisfy the
requirement that the W-matrix for particle and anti-particle coincide
(\ref{eq: antip}), we may simply use the fact that $\lambda_i \cdot \sigma^q
\gamma_j = \lambda_j \cdot \sigma^q \gamma_i$ \cite{FO}. Meromorphicity
follows from the same argument employed in \cite{FO} for the S-matrix.
Because of relation (\ref{eq: merom}), each block occurs twice in the product,
where the values of $q$ are related by $ q + q' = h + \frac{c(i)+c(\ib)}{2}$.
Then the total power of each block turns out to be $ - \lambda_i \cdot \sigma^q
\gamma_{\ib}$, which for simply laced algebras will always be an integer.
\par
Next we verify (\ref{eq: crossunw}). We have
\beq
W_{\ib}(\th + i \pi) W_i(\th)= \prod_{q=a}^{h-1+a} \left( {\cal W}_{2q-
\frac{c(i)+ c(\ib)}{2}}(\th) {\cal W}_{2q +2h - \frac{c(i)+c(\ib)}{2}}(\th)
 \right) ^ {- \frac{1}{2}\lambda_i \cdot \sigma^q \gamma_{\ib} } \;\; .
\eeq
Although $W_x(\th)$ is not $2 h$-periodic in $x$, the expression in the bracket
of this equation is. Together with the fact that the Coxeter element has
period $h$, we are permitted to shift the dummy variable q by
\beq
q \rightarrow  q - \frac{h}{2} + \frac{ c(\ib) - c(i)}{4} \;\; .
\eeq
Then employing (\ref{eq: ssx}) and the relation between simple roots associated
to particle and  antiparticle $ \gamma_{\ib} = - \sigma^{ - \frac{h}{2} +
\frac{c(i)+c(\ib)}{4}} \gamma_i $ \cite{FO}, the expression becomes
\beq
\prod_{q=1}^{h} \left\{ 2q - c(i) \right\}_{2 \th}^{-\frac{1}{2} \lambda_i
\cdot \sigma^q \gamma_{i} } \; =\; S_{ii}(2 \th)  \;\;
\eeq
and hence (\ref{eq: wmat}) solves the ``crossing unitarity" relation.
\par

In order to establish that this function satisfies the wall-boootstrap equation
(\ref{eq:bootw}), we would like to use a similar trick as in
the case of the S-matrix:  we would like to be able to shift the dummy
variable q. However, since ${\cal W}_{2q + \hbox{const}}(\th)$ has period $2h$
in q and $ \sigma^q$ has period h, a shift in q by some integer value will
alter
the expression for $W_i(\th)$. In fact it turns out that this expression does
no
   t solve the wall-bootstrap equation. However, the above arguments exhibit
the
the reason for the following assumption:
\par
 We presume that the blocks $\wc_x$, which constitute the W-matrix are,
{\em up to a shift of $2h$ in $x$}, in one-to-one correspondence to the blocks
$ \{ x \}_{\th}$ which build up the S-matrix. Additional factors can only
be CDD-ambiguities, which we will as usual ignore. The problem which remains is
   to determine which of the blocks are shifted by $2h$ and which are not. In
ot
   her words, the bootstrap equation (\ref{eq:bootw}) will determine whether
the
    block
$\{ x\}$ in the S-matrix is to be replaced by ${\cal W}_x(\th)$ or  ${\cal
W}_{x
   +2h}(\th)$.
\par
At present we do not have a general unified argument valid for all Toda
theories at hand and we shall therefore turn to a case-by-case analysis.
The W-matrix for all
Toda theories related to simply laced Lie algebras will now be obtained in
following way: Writing down first the whole set of W-bootstrap equations, with
t
   he fusing angles  for instance obtained from \cite{BCDS} or an explicit
computation of the fusing rules, we seek equations involving only one
particular $W_i(\th)$. We then find the most general solution of this equation.
Having found one $W_i(\th)$ we seek an equation involving one additional
$W_j(\th)$, which then is computed. Proceeding in this fashion we are able
to construct $W_i(\th)$ for $i=1, \dots, r$. The remaining equations then have
to be satisfied identically. In this manner we obtain the following
solutions:

\subsection{ ${\bf a_n^{(1)} } $}

Using as convention that $c(1)$ is always $-1$, we derive for $i=1, \dots, r$
\beq
W_i(\th) \; = \; \prod_{l=1}^{\mu(i)} {\cal{W}}_{ r + 2 \nu(l) - 2 \mu(i)}
(\th) \;\; .
\eeq
Here $\mu(i)$ is a function which takes the $Z \!\!\!Z_2$-symmetry of the
Dynkin diagram into account, i.e.
\beq
\mu(i) := \left\{
\begin{array}{ll}
i   & \qquad \hbox{for} \;  i \leq [\frac{h}{2}] \\
h-i & \qquad \hbox{for} \;  i > [\frac{h}{2}]  \;\; ,
\end{array}    \right.
\eeq
and $\nu$ is a function defined as
\beq
\nu(n) := \left\{
\begin{array}{ll}
n   & \qquad \hbox{for} \;  n \; \hbox{odd} \\
n+h & \qquad \hbox{for} \;  n \; \hbox{even}  \;\; .
\end{array}    \right.
\eeq
At present we do not know a solid mathematical proof for this formula
to any order, but we have checked its validity to high order in $n$.

\subsection{ ${\bf d_n^{(1)} }$ }

In this case we adopt the same convention $c(1)=-1$, such that we obtain for
the two ``spinors" $c(r) = c(r-1) = 1$ or $c(r) = c(r-1) = -1$ ,
depending on whether $r$ is odd or even, respectively. Then we obtain
for $i=1, \dots, r-2$
\bea
W_i(\th) & = & \prod_{l=1}^{i} {\cal{W}}_{ 2i -2l +1}(\th) \;
                    {\cal{W}}_{h - 2i +2 \nu(l) -1}(\th) \\
W_r(\th) & = &  W_{r-1}(\th) = \prod_{l=1}^{[\frac{r}{2}]}
{\cal{W}}_{4l-3}(\th)  \;{\cal{W}}_{h-4l+3}(\th)
\eea
Again some inductive proof is still required.

\subsection{ ${\bf e_6^{(1)} }$ }

In order to avoid cumbersome expressions we introduce the following
symbol
\beq
_l\!\left[{\cal W}_x(\th) \right]_{_m}^y \;\equiv  \;
{\cal W}_x^{y-l}(\th)\; {\cal W}_{h-x}^{y-m}(\th)\; {\cal W}_{x+2h}^{l}(\th)
\;{\cal W}_{3h-x}^{m}(\th)
\eeq
Our conventions are illustrated in the following Dynkin diagram

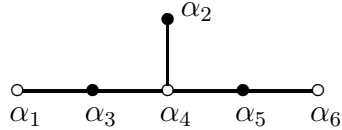
\setlength{\unitlength}{0.01cm}
\begin{picture}( 1000,300)(0,100)
\thicklines
\put(492,190){$ \circ$}
\put(510,200){\line( 1, 0){85}}
\put(592,190){$ \bullet$}
\put(702,207){\line( 0, 1){85}}
\put(692,285){$ \bullet$}
\put(610,200){\line( 1, 0){85}}
\put(692,190){$ \circ$}
\put(792,190){$ \bullet$}
\put(892,190){$ \circ$}
\put(710,200){\line( 1, 0){85}}
\put(810,200){\line( 1, 0){85}}
\put(492,160){$ \alpha_1$}
\put(592,160){$ \alpha_3$}
\put(692,160){$ \alpha_4$}
\put(792,160){$ \alpha_5$}
\put(892,160){$ \alpha_6$}
\put(720,300){$ \alpha_2$}
\end{picture}

We then derive
\bea
W_1(\th) & =& W_6(\th) \;=\; {\cal{W}}_{5}(\th)\; {\cal{W}}_{35}(\th)  \\
W_2(\th) & = & \left[{\cal{W}}_{1}(\th) \right]^1 \; _1\! \left[{\cal{W}
                  }_{5}(\th) \right]^1  \\
W_3(\th) & = & W_5(\th) \;=\; {\cal{W}}_{3}(\th)\; {\cal{W}}_{5} (\th)\;
           {\cal{W}}_{7}(\th)\; {\cal{W}}_{11}(\th)\;{\cal{W}}_{29}(\th)
            \;{\cal{W}}_{33}(\th)  \\
W_4(\th) & = &  \left[{\cal{W}}_{1}(\th) \right]_{_1}^1 \;
               _1\!\left[{\cal{W}}_{3}(\th) \right]^2 \;
               _1\!\left[{\cal{W}}_{5}(\th) \right]_{2}^3  \;\; .
\eea

\subsection{ ${\bf e_7^{(1)} }$ }

Again we depict our conventions in a Dynkin diagram

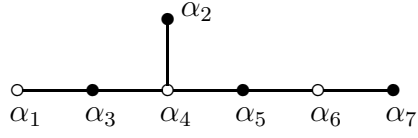
\setlength{\unitlength}{0.01cm}
\begin{picture}( 1000,300)(0,100)
\thicklines
\put(492,190){$ \circ$}
\put(510,200){\line( 1, 0){85}}
\put(592,190){$ \bullet$}
\put(702,207){\line( 0, 1){85}}
\put(692,285){$ \bullet$}
\put(610,200){\line( 1, 0){85}}
\put(692,190){$ \circ$}
\put(792,190){$ \bullet$}
\put(892,190){$ \circ$}
\put(710,200){\line( 1, 0){85}}
\put(810,200){\line( 1, 0){85}}
\put(910,200){\line( 1, 0){85}}
\put(992,190){$ \bullet$}
\put(992,160){$ \alpha_7$}
\put(492,160){$ \alpha_1$}
\put(592,160){$ \alpha_3$}
\put(692,160){$ \alpha_4$}
\put(792,160){$ \alpha_5$}
\put(892,160){$ \alpha_6$}
\put(720,300){$ \alpha_2$}
\end{picture}

and obtain

\bea
W_1(\th) & = & \left[{\cal{W}}_{1}(\th) \right]_{1}^1 \;
\left[{\
   cal{W}}_{7}(\th) \right]^1 \\
W_2(\th) & = & \left[{\cal{W}}_{1}(\th) \right]^1 \;
                \left[{\cal{W}}_{5}(\th) \right]^1 \;
            \left[{\cal{W}}_{7}(\th) \right]_{1}^1 \; {\cal{W}}_{9}(\th)  \\
W_3(\th) & = & \left[{\cal{W}}_{1}(\th) \right]^1 \;
   \left[{\cal{W}}_{3}(\th)\right]_{1}^1\;\left[{\cal{W}}_{5}(\th)\right]^1
                \; _1\!\left[{\cal{W}}_{7}(\th)\right]_{1}^2
                           \;\left[{\cal{W}}_{9}(\th)\right]^1 \\
W_4(\th) & = & \left[{\cal{W}}_{1}(\th) \right]^1_{1} \;
   \left[{\cal{W}}_{3}(\th)\right]^2\;_{1} \!\left[{\cal{W}}_{5}(\th)
     \right]_{_2}^3 \; _1\!\left[{\cal{W}}_{7}(\th)\right]_{1}^4
     \; _1\!\left[{\cal{W}}_{9}(\th) \right]_{1}^2 \\
W_5(\th) & = & \left[{\cal{W}}_{1}(\th) \right]^1 \;
   \left[{\cal{W}}_{3}(\th)\right]^1_{1}\;\left[{\cal{W}}_{5}(\th)\right]^2 \;
   _1\!\left[{\cal{W}}_{7}(\th)\right]_{1}^2\;\left[{\cal{W}}_{9}(\th)
        \right]_{1}^1 \; {\cal{W}}_{9}(\th)   \\
W_6(\th) & = & \left[{\cal{W}}_{1}(\th) \right]^1_{1} \;
   \left[{\cal{W}}_{3}(\th)\right]^1\; \left[{\cal{W}}_{7}(\th)\right]^1 \;
   \left[{\cal{W}}_{9}(\th)\right]_{1}^1   \\
W_7(\th) & = & \left[{\cal{W}}_{1}(\th) \right]^1 \; {\cal{W}}_{9}(\th) \;\; .
\eea

\subsection{ ${\bf e_8^{(1)} }$ }

Together with the notations

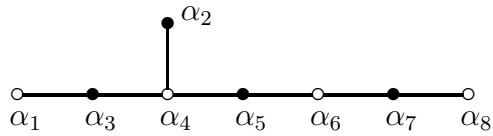
\setlength{\unitlength}{0.01cm}
\begin{picture}( 1100,300)(0,100)
\thicklines
\put(492,190){$ \circ$}
\put(510,200){\line( 1, 0){85}}
\put(592,190){$ \bullet$}
\put(702,207){\line( 0, 1){85}}
\put(692,285){$ \bullet$}
\put(610,200){\line( 1, 0){85}}
\put(692,190){$ \circ$}
\put(792,190){$ \bullet$}
\put(892,190){$ \circ$}
\put(710,200){\line( 1, 0){85}}
\put(810,200){\line( 1, 0){85}}
\put(910,200){\line( 1, 0){85}}
\put(992,190){$ \bullet$}
\put(992,160){$ \alpha_7$}
\put(1010,200){\line( 1, 0){85}}
\put(1092,190){$ \circ$}
\put(1092,160){$ \alpha_8$}

\put(492,160){$ \alpha_1$}
\put(592,160){$ \alpha_3$}
\put(692,160){$ \alpha_4$}
\put(792,160){$ \alpha_5$}
\put(892,160){$ \alpha_6$}
\put(720,300){$ \alpha_2$}
\end{picture}

we obtain
\bea
W_1(\th) & = & \left[{\cal{W}}_{1}(\th) \right]^1_1 \;
   \left[{\cal{W}}_{7}(\th)\right]^1\;\left[{\cal{W}}_{11}(\th)\right]^1 \;
   \left[{\cal{W}}_{13}(\th) \right]_1^1   \\
W_2(\th) & = & \left[{\cal{W}}_{1}(\th) \right]^1 \;
   \left[{\cal{W}}_{5}(\th)\right]^1\;\left[{\cal{W}}_{7}(\th)\right]^1_1 \;
\left[{\cal{W}}_{9}(\th)\right]^1\;_1\!\left[{\cal{W}}_{11}(\th)\right]^2_1\;
    \left[{\cal{W}}_{13}(\th) \right]^1  \nn \\
   & & \left[{\cal{W}}_{15}(\th)\right]_1^1 \\
W_3(\th) & = & \left[{\cal{W}}_{1}(\th) \right]^1 \;
   \left[{\cal{W}}_{3}(\th)\right]^1_1\;\left[{\cal{W}}_{5}(\th)\right]^1\;
  _1\!\left[{\cal{W}}_{7}(\th)\right]^2_1\;\left[{\cal{W}}_{9}(\th)\right]^2\;
  _1\!\left[{\cal{W}}_{11}(\th) \right]^3_2  \nn \\
   & &   _1\!\left[{\cal{W}}_{13}(\th) \right]_1^3 \; \left[{\cal{W}}_{15}(\th)
        \right]^1_1    \\
W_4(\th) & = & \left[{\cal{W}}_{1}(\th) \right]^1_1 \;
   \left[{\cal{W}}_{3}(\th)\right]^2 \;_1\!\left[{\cal{W}}_{5}(\th)
    \right]^3_2\;  _1\!\left[{\cal{W}}_{7}(\th)\right]^4_1  \nn \\
    & &   _2\!\left[{\cal{W}}_{9}(\th)\right]^5_3\;
        _2\!\left[{\cal{W}}_{11}(\th) \right]^6_2 \;
       _3\!\left[{\cal{W}}_{13}(\th) \right]_3^6  \;
         \left[ {\cal{W}}_{15}(\th) \right]^3_2   \\
W_5(\th) & = & \left[{\cal{W}}_{1}(\th) \right]^1 \;
   \left[{\cal{W}}_{3}(\th)\right]^1_1 \;\left[{\cal{W}}_{5}(\th)\right]^2\;
  _1\!\left[{\cal{W}}_{7}(\th)\right]^3_2\;_1\!\left[{\cal{W}}_{9}(\th)
   \right]^3_1\; _2\!\left[{\cal{W}}_{11}(\th) \right]^4_2 \nn  \\
  & &   _1\!\left[{\cal{W}}_{13}(\th)
    \right]_1^4  \; \left[{\cal{W}}_{15}(\th) \right]^2_2   \\
W_6(\th) & = & \left[{\cal{W}}_{1}(\th) \right]^1_1 \;
   \left[{\cal{W}}_{3}(\th)\right]^1 \;\left[{\cal{W}}_{5}(\th)\right]^1_1\;
  \left[{\cal{W}}_{7}(\th)\right]^1 \;_1\!\left[{\cal{W}}_{9}(\th)\right]^2_1\;
  _1\!\left[{\cal{W}}_{11}(\th) \right]^3_1  \nn \\
  & & _1\!\left[{\cal{W}}_{13}(\th)
    \right]_1^2  \; \left[{\cal{W}}_{15}(\th) \right]^1  \\
W_7(\th) & = & \left[{\cal{W}}_{1}(\th) \right]^1 \;
   \left[{\cal{W}}_{3}(\th)\right]^1_1 \; \left[{\cal{W}}_{9}(\th)\right]^1\;
  _1\!\left[{\cal{W}}_{11}(\th) \right]^2_1\;_1\!\left[{\cal{W}}_{13}(\th)
    \right]^1   \\
W_8(\th) & = & \left[{\cal{W}}_{1}(\th) \right]^1_1 \;
     \left[{\cal{W}}_{11}(\th) \right]^1 \;\; .
\eea
We have here reported only one solution. As explained above a second solution
can always be obtained by multiplication of the CDD-factor or equivalently
by shifting all the xs by $2h$ in $\wc_x$. Both relations will give rise to
entirely different physics due to the different positions of the poles in
the physical sheet. Depending on the order and the sign of the residue
several of these states may find an interpretation as stable states in the
bound
   ary.

\section{Conclusion}

We have demonstrated how the set of consistency equations for the scattering
matrices in the presence of reflecting boundaries can be employed in order
to compute the W-matrices for affine Toda field theories.
\par
Evidently the completion of the picture requires further investigations and
several interesting questions have still been left unanswered. Concerning
ATFTs,
    a detailed study of possible integrable boundary conditions is
desirable, which might illuminate further their relation to integrable
deformed conformal field theories. It is also expected that the W-matrix
can be cast into a more general unified formula, analogous to the S-matrix,
which might lead to a deeper Lie algebraic understanding. Since most of
the solutions for the W-matrix exhibit the possibility of stable bound states,
we may relax our assumption that the wall is always in ground state and
determine the matrices which incorporate these states.
Finally the question, of whether it is possible to parallel the argumentation
in the case of the absence of boundaries and use the knowledge obtained
on-shell in order to determine off-shell properties of the theory, poses
an interesting problem.

\vfill \break
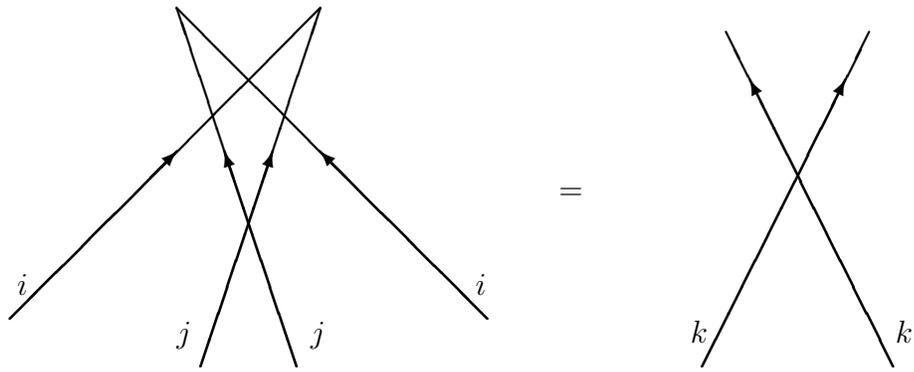
\begin{figure}
\setlength{\unitlength}{0.0125in}
\begin{picture}(40,0)(60,470)
\thicklines
\put(230,500){\line( 1, 3){50}}
\put(270,500){\line( -1, 3){50}}
\put(280,650){\line(-1,-1){130}}
\put(220,650){\line(1,-1){130}}
\put(230,500){\vector( 1, 3){30}}
\put(270,500){\vector( -1, 3){30}}
\put(150,520){\vector( 1, 1){70}}
\put(350,520){\vector(- 1, 1){70}}
\put(440,500){\line( 1, 2){70}}
\put(520,500){\line( - 1, 2){70}}
\put(440,500){\vector( 1, 2){60}}
\put(520,500){\vector(- 1, 2){60}}
\put(380,570){$ =$}
\put(220,510){$ j$}
\put(276,510){$ j$}
\put(153,530){$ i$}
\put(345,530){$ i$}
\put(435,510){$ k$}
\put(521,510){$ k$}
\end{picture}
 \caption{The bootstrap equation (2.10) }
 \end{figure}
\begin{figure}
\setlength{\unitlength}{0.0125in}
\begin{picture}(40,90)(60,420)
\thicklines
\put(120,510){\line( 1,0){170}}
\put(120,500){\line( 1,0){170}}
\put(120,500){\line( 0,1){10}}
\put(290,500){\line( 0,1){10}}
\put(120,500){\line(2,1){20}}
\put(130,500){\line(2,1){20}}
\put(140,500){\line(2,1){20}}
\put(150,500){\line(2,1){20}}
\put(160,500){\line(2,1){20}}
\put(170,500){\line(2,1){20}}
\put(180,500){\line(2,1){20}}
\put(190,500){\line(2,1){20}}
\put(200,500){\line(2,1){20}}
\put(210,500){\line(2,1){20}}
\put(220,500){\line(2,1){20}}
\put(230,500){\line(2,1){20}}
\put(240,500){\line(2,1){20}}
\put(250,500){\line(2,1){20}}
\put(260,500){\line(2,1){20}}
\put(270,500){\line(2,1){20}}
\put(200,510){\line(-3,2){72}}
\put(200,510){\line(3,2){72}}
\put(128,558){\vector(3,-2){36}}
\put(200,510){\vector(3,2){36}}
\put(111,550){$i$}
\put(200,500){\line(-3,-2){72}}
\put(200,500){\line(3,-2){72}}
\put(128,452){\vector(3,2){36}}
\put(200,500){\vector(3,-2){36}}
\put(111,550){$i$}
\put(170,515){$ \theta$}
\put(350,558){\vector(3,-2){159}}
\put(350,452){\vector(3,2){159}}
\put(405,500){$ 2\theta$}
\put(225,485){$ \theta + i \pi$}
\put(340,558){$i$}
\put(340,447){$i$}
\put(310,505){$=$}
\end{picture}
\caption{The ``crossing unitarity" relation}
 \end{figure}
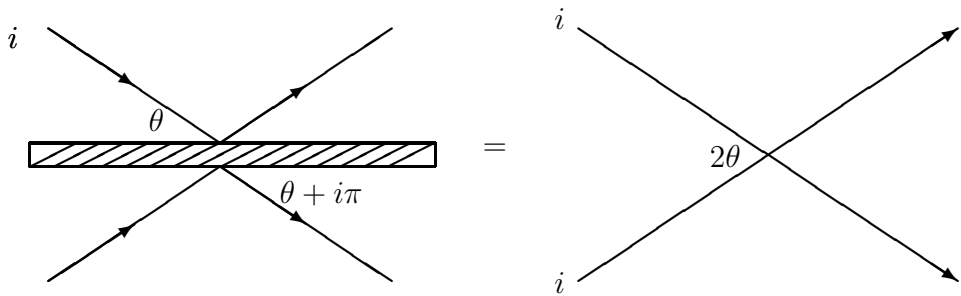

\end{document}